\documentstyle[aps,prb,preprint,graphicx]{revtex}

\begin{document}

\draft

\title{Supersymmetric Mean-Field Theory of $t$-$J$ Model}

\author{C.H.Cheng and T.K.Ng}
\address{Department of Physics, Hong Kong University of Science and
Technology, Clear Water Bay, Hong Kong, China}

\date{Received \today}
		
\maketitle

\begin{abstract}
The supersymmetric formulation of $t$-$J$ model is studied in this paper at the mean-field level 
where the $\delta$-$T$ phase diagram is computed. We find that slave-fermion-like spiral phase 
is stable at low doping concentration, and the slave-boson-like d-wave fermionic spin pairing state 
becomes energetically favourable when $\delta\geq 0.23$. An improvement in free energy using 
Gutzwiller's method lowers the transition doping concentration to $\delta\sim 0.06$.
We also point out the existence of new branches of excitations in the supersymmetric theory.
\end{abstract}


\narrowtext

A lot of interest in studying the $t$-$J$ model is raised since the discovery of cuprate 
superconductors\cite{and}. So far,
analytical understandings of the model were largely based on mean-field theories which treat 
the constraint of no double occupany on average. The most successful mean-field approaches to 
the $t$-$J$ model seem to be based on either the slave-fermion mean-field theory (SFMFT)\cite{sign}, 
which is successful at very small doping when antiferromagnetic correlation is important, 
or the slave-boson mean-field theory (SBMFT)\cite{sb}, which is successful at larger value 
of doping when the system becomes superconducting. The only difference between the two approaches 
is that two different representations of spin and hole operators are used in the two theories.
Recently the focus has turned to the underdoped regime where crossover from 
antiferromagnetism to superconductivity takes place \cite{theory1,theory2}. 
To understand the complicated behaviour 
in this regime, it seems that a unified approach which incorporates the advantages of both SFMFT 
and SBMFT is essential. In a recent paper\cite{super}, we show that it is in general possible to 
formulate lattice models with constraints of no double occupancy in such a way that the 
advantages of the two slave-particle representations are retained. 
The essential idea of the approach is to consider 
an enlarged Hilbert space where both slave-boson and slave-fermion representation of spins and 
holes coexist as {\it independent} states at each lattice site. A Hamiltonian is chosen such that 
the system in the enlarged Hilbert space is equivalent to $2^N$ ($N=$ number of lattice sites) 
replicas of the original lattice model (supersymmetric formulation)\cite{super}. 

In this paper, we present our mean-field calculation results on the supersymmetric formulation 
of $t$-$J$ model. First we give some mathematical details of the theory. A spin-$\sigma$ 
state is represented by either $b^{\dag}_{\sigma}|0\rangle$ or $f^{\dag}_{\sigma}|0\rangle$, 
whereas a hole is represented by either ${\cal F}^{\dag}|0\rangle$ or ${\cal B}^{\dag}|0\rangle$, 
where $b_{\sigma}^{\dag}$, ${\cal B}^{\dag}$ are bosonic creation operators and  
$f_{\sigma}^{\dag}$, ${\cal F}^{\dag}$ are fermionic ones. 
The Hamiltonian which is equivalent to $2^N$ replicas of $t$-$J$ model in the 
Hilbert space spanned by the above states is\cite{super}
\begin{eqnarray}
\label{ham}
H&=&-t\sum_{\langle i,j\rangle,\sigma}(
-b_{i\sigma}^{\dag}{\cal{F}}_{i}{\cal{F}}_{j}^{\dag}b_{j\sigma}
+b_{i\sigma}^{\dag}{\cal{F}}_{i}{\cal{B}}_{j}^{\dag}f_{j\sigma}
+f_{i\sigma}^{\dag}{\cal{B}}_{i}{\cal{F}}_{j}^{\dag}b_{j\sigma} 
+f_{i\sigma}^{\dag}{\cal{B}}_{i}{\cal{B}}_{j}^{\dag}f_{j\sigma} + h.c.) \nonumber \\ 
&&+J\sum_{\langle i,j\rangle}{{\bf S}_i\cdot{\bf S}_j} 
+\sum_i\lambda_i(h_i^{\dag}h_i
+\sum_{\sigma}\psi_{i\sigma}^{\dag}\psi_{i\sigma}-1)
-\mu\sum_i h_i^{\dag}h_i
\end{eqnarray}
where
\begin{eqnarray}
{\bf S}_i=\frac{1}{2}
\left(\begin{array}{cc}b_{i\uparrow}^{\dag} & b_{i\downarrow}^{\dag} 
\end{array}\right)      
\vec{\sigma}
\left(\begin{array}{c}b_{i\uparrow} \\ b_{i\downarrow}
\end{array}\right)
+\frac{1}{2}
\left(\begin{array}{cc}f_{i\uparrow}^{\dag} & f_{i\downarrow}^{\dag} 
\end{array}\right)      
\vec{\sigma}
\left(\begin{array}{c}f_{i\uparrow} \\ f_{i\downarrow}
\end{array}\right),
\end{eqnarray}
$\vec{\sigma}$ is the Pauli matrix,
and 
\begin{equation}
h_i=\left(\begin{array}{c}{\cal F}_i \\ {\cal B}_i \end{array}\right),
\psi_{i\sigma}=\left(\begin{array}{c}b_{i\sigma} \\ f_{i\sigma}
\end{array}\right).
\end{equation}
The constraint of no double occupancy 
$h_i^{\dag}h_i+\sum_{\sigma}\psi_{i\sigma}^{\dag}\psi_{i\sigma}=1$
is imposed by a Lagrange multiplier term as usual.
The Heisenberg interaction can be re-written as 
\begin{equation}
\label{hj}
H_J=-\frac{J}{4}\sum_{\langle i,j\rangle}
{\rm str}(\Lambda
\Delta_{ij}^{\dag}\Delta_{ij}+\chi_{ij}^{\dag}\chi_{ij})
\end{equation}
where 
\begin{equation}
\Delta_{ij}=\left(\begin{array}{cc}
\Delta_{ij}^{bb} & \Delta_{ij}^{bf} \\ \Delta_{ij}^{fb} & \Delta_{ij}^{ff}
\end{array}\right),
\chi_{ij}=\left(\begin{array}{cc}
\chi_{ij}^{bb} & \chi_{ij}^{bf} \\ \chi_{ij}^{fb} & \chi_{ij}^{ff}
\end{array}\right),
\Lambda=\left(\begin{array}{cc}1 & 0 \\ 0 & -1\end{array}\right).
\end{equation}
$\Delta_{ij}^{\alpha\beta}=\alpha_{i\uparrow}\beta_{j\downarrow}-\alpha_{i\downarrow}\beta_{j\uparrow}$, $\chi_{ij}^{\alpha\beta}
=\sum_{\sigma}\alpha_{i\sigma}^{\dag}\beta_{j\sigma}$, where
$\alpha_{i\sigma},\beta_{i\sigma}$ can be either $b_{i\sigma}$ or $f_{i\sigma}$. Notice that 
only $\Delta^{bb(ff)}_{ij}$ and $\chi^{bb(ff)}_{ij}$ terms appear in usual slave-fermion(boson) 
mean-field theory, whereas $\Delta_{ij}$ and $\chi_{ij}$ are supermatrices in our formulation. 
It is straightforward to show from Eq.\ (\ref{hj}) that $H_J$ is invariant under the super-unitary 
transformation 
$h_i\rightarrow U_i h_i, \psi_{\sigma}\rightarrow U_i\psi_{\sigma}$,  
where $U_i$ is a local $2\times2$ unitary super-matrix. The hopping term can also be re-written as
\begin{equation}
H_t=-t\sum_{\langle i,j\rangle}{\rm str}({\cal{H}}_{ij}^{\dag}\chi_{ij}+h.c.),
\end{equation}
where 
\begin{equation}
{\cal H}_{ij}=\left(\begin{array}{cc}
-{\cal{F}}_{j}{\cal{F}}_{i}^{\dag} & {\cal{F}}_{j}{\cal{B}}_{i}^{\dag} \\
{\cal{B}}_{j}{\cal{F}}_{i}^{\dag} & {\cal{B}}_{j}{\cal{B}}_{i}^{\dag}
\end{array}\right).	
\end{equation}
Notice that $H_t$ is not invariant under the super-unitary 
transformation $U_i$ because of the sign difference in the term
$-b_{i\sigma}^{\dag}{\cal{F}}_{i}{\cal{F}}_{j}^{\dag}b_{j\sigma}$ 
in Eq.(\ref{ham}). The sign difference is well known in SFMFT \cite{sign}. 
Thus although the 
Heisenberg model is supersymmetric, the $t$-term breaks supersymmetry.
The Lagrangian ${\cal L}$ of $t$-$J$ model in the enlarged Hilbert space is 
${\cal L}=\sum_{i,\sigma}\psi_{i\sigma}^{\dag}\partial_{\tau}\psi_{i\sigma}
+\sum_i h_i^{\dag}\partial_{\tau}h_i+H$.

Our mean-field theory is obtained by replacing $\lambda_i\rightarrow\langle\lambda_i\rangle=\lambda$,
and by decoupling the quartic terms appearing in 
$H_t$ and $H_J$. 
After some straightforward algebra, we obtain
\begin{eqnarray}
{\cal L}&\rightarrow&\sum_{i,\sigma}\psi_{i\sigma}^*\partial_{\tau}\psi_{i\sigma}
+\sum_i h_i^*(\partial_{\tau}-\mu)h_i
+\sum_i\lambda_i(h_i^*h_i+\sum_{\sigma}\psi_{i\sigma}^*\psi_{i\sigma}-1) \nonumber \\
&&-t\sum_{\langle i,j\rangle}{\rm str}(\tilde{\cal H}^*_{ij}\chi_{ij}+{\cal H}^*_{ij}\tilde{\chi}_{ij}-\tilde{\cal H}_{ij}\tilde{\chi}_{ij} + c.c.)
-\frac{J}{4}\sum_{\langle i,j\rangle} {\rm str}(\tilde{\chi}_{ij}^*\chi_{ij}+\chi_{ij}^*\tilde{\chi}_{ij}-\tilde{\chi}_{ij}^*\tilde{\chi}_{ij}) \nonumber \\
&&-\frac{J}{4}\sum_{\langle i,j\rangle} {\rm str} \Lambda
(\tilde{\Delta}_{ij}^*\Delta_{ij}+\Delta_{ij}^*\tilde{\Delta}_{ij}-\tilde{\Delta}_{ij}^*\tilde{\Delta}_{ij})
\end{eqnarray}
where $\tilde{\Delta}_{ij} (\tilde{\Delta}_{ij}^*)$, $\tilde{\chi}_{ij} (\tilde{\chi}_{ij}^*)$ and
$\tilde{\cal H}_{ij} (\tilde{\cal H}_{ij}^*)$ are Hubbard-Stratonvich type super-matrix field
introduced to decouple the quartic terms in $H$.

Next we integrate out hole and spin fields $h$ and $\psi_{\sigma}$, arriving in an 
action in terms of the auxiliary super-matrix fields only. 
The supersymmetric mean-field theory is obtained by replacing the auxiliary 
super-matrix fields by their mean values which are determined 
by the saddle-point conditions,
\begin{equation}
\frac{\delta {\cal S}}{\delta Q_{ij}(\tau)}=
\frac{\delta {\cal S}}{\delta Q_{ij}^{*}(\tau)}=
\frac{\delta {\cal S}}{\delta \lambda_i(\tau)}=0
\end{equation}
where $Q_{ij}$ can be $\tilde{\Delta}_{ij}^{\alpha\beta}$,
$\tilde{\chi}_{ij}^{\alpha\beta}$ or $\tilde{\cal H}_{ij}^{\alpha\beta}$.
Since the first-derivative of an action $S$ with respect to any Grassmann
field is a Grassmann variable and mean-field value of any Grassmann variable
is zero, all the off-diagonal elements of the super-matrices vanish at the mean-field level 
and only diagonal terms of the supermatrix fields survive. 
Notice that both slave-boson and slave-fermion like mean-field parameters 
are present in the mean-field Hamiltonian. The relative weight of the two kinds of
terms are determined by minimizing the free energy of the system with the 
constraint that the correct {\it average} number of spins and holes are obtained \cite{super}.

In our calculation we have only searched for mean-field solutions which preserve 
translational invariance and time-reversal symmetry. 
Notice that the ground state energy of the Heisenberg term
obtained from the corresponding variational wavefunction method
\cite{gutz2,gutz3} is 
$E^{var}_J=2 E^{mf}_J$,
where $E^{mf}_J$ is the energy in our mean-field theory. 
This discrepancy is due to the defect of Hubbard-Stratonovich 
method when decoupling in more than one channel \cite{lee}. 
To compare our results with those obtained from
variational wavefunction method, the coupling constant $J$ in our mean-field theory
is scaled to $2J$ in the following. 

Fig.1 shows the $\delta$-$T$ phase diagram with $t/J=3$.
The phase diagram can be divided into four regions.
There are three low temperature phases at different values of $\delta$
and a high temperature phase (IV) where all order parameters vanish. At low doping 
the stable low temperature phase is slave-fermion like (I), with the system characterized by 
antiferromagnetic or spiral order, whereas at higher doping the stable low temperature phase is 
slave-boson like, with the system characterized by d-wave RVB pairing (II) or is 
a Fermi liquid (III) (at high doping). The phase transitions are all first-order (solid lines)
except that between d-wave fermionic RVB and Fermi liquid (dot-dashed line) which is second-order.
Notice that first-order transition to high-temperature phase is also obtained in pure
slave-boson or slave-fermion treatment, and is believed to be a defect
of mean-field theory. The same is also expected here. 
It is however, not so clear whether the first-order transition from slave-fermion-like
to slave-boson-like solution is a defect of mean-field theory, or whether 
it really exists in {\it t-J} model. 
It is interesting to point out the similarity of our mean-field phase diagram to that
obtained from SO(5) theory \cite{theory2}. 
Assuming that the first-order phase transition to high-temperature phases are replaced
by smooth crossover,
the two phase diagram are qualitatively 
similar, except the presence of spin-charge separation and the replacement of 
antiferromagnetic phase by spiral phase in our mean-field theory. Whether there exists
a hidden SO(5) symmetry in our theory is not clear at present. 
In the following,
we shall describe the structure of the phase diagram in more details.

At half-filling, antiferromagnetic state with only $\langle\Delta_{ij}^{bb}\rangle\neq0$ is the 
most stable mean-field solution until at sufficient high temperature $T\geq0.6J$, 
where $\langle \Delta_{ij}^{bb}\rangle$ vanishes and the system is in the high temperature 
phase similar to what happen in the usual SFMFT. For $0<\delta\leq 0.23$, 
$\langle\chi_{ij}^{bb}\rangle$ 
becomes non-zero at low temperature, corresponding to the development of spiral order. 
Upon further doping the spiral phase becomes energetically unfavorable and is replaced  
via a first-order transition by
the slave-boson-like d-wave RVB pairing phase where only $\langle\Delta_{ij}^{ff}\rangle$ 
and $\langle\chi_{ij}^{ff}\rangle$ are nonzero. At still larger doping or higher temperature, 
the slave-boson-like d-wave RVB pairing phase is replaced by the Fermi liquid and 
``strange metal'' phases, respectively with only $\langle\chi_{ij}^{ff}\rangle\neq 0$. 
Notice also that as in the usual SBMFT, a nonzero d-wave RVB pairing order parameter 
$\langle\Delta_{ij}^{ff}\rangle$ may describe either superconducting or spin-gap phases, 
depending on whether the holons are bose-condensed. The bose-condensation temperature $T_B$ 
is found to be very high in supersymmetric mean-field theory as in the usual SBMFT (dashed line). 
$T_B$ is believed to be suppressed by gauge fluctuations once we go beyond 
mean-field theory\cite{bose}. We have also searched for "mixed" states where both slave-fermion 
and slave-boson type order parameters coexist. However we found numerically that such phases 
are never the most stable. The most favorable phase is always characterised by slave-fermion-like 
or slave-boson-like order parameters except at high enough temperature where all the order 
parameters vanish. 

Notice that although the mean-field solutions we obtained are always either slave-boson-like or 
slave-fermion-like, there is an important difference between supersymmetric theory and 
usual slave-fermion or slave-boson approaches. In usual slave-fermion or slave-boson 
mean-field theories the statistics of the spin and hole excitations are fixed and are 
determined by the types of mean-field theory being used. However, in the supersymmetric 
theory spin and hole excitations with {\em both} fermi- and bose- statistics exist \cite{super}! 
For example, in the spiral phase 
where the ground state is the same as that obtained in the usual slave-fermion 
mean-field theory, we find that besides the usual bosonic spin-wave excitation, new dispersionless
high energy 
fermionic spin excitation of energy $\sim 2.3J$ and bosonic hole excitation of
energy $\sim 10\delta t$ appears in the excitation spectrum. 
The fermionic spin excitation energy decreases as the 
antiferromagnetic correlation weakens either because of raising temperature or increasing doping 
concentration, until the slave-boson-like d-wave RVB pairing phase 
becomes energetically more stable. The excitation spectrum in the slave-boson-like phase 
has a "dual" structure to the antiferromagnetic phase. The low energy spin excitations are 
fermionic as in the usual SBMFT. However, a dispersionless high energy bosonic spin excitation 
of energy $\sim 0.4J$ and fermionic hole excitation of energy $\sim 1.5t$ (at low doping regime)
are found to exist in the supersymmetric theory. 
Notice that finite dispersion is expected to appear in these high energy excitations
after the fluctuation arising from the off-diagonal elements of super-matrices is considered. 

One of us proposed that these high energy spin and charge excitations are in fact topological 
excitations in SFMFT and SBMFT, respectively \cite{ng1,ng2}. It is suggested 
that the slave-fermion (boson) mean-field state can be considered as a topologically-disordered 
state of slave-boson (fermion) mean-field theory. In the transition from one mean-field 
state to another, the (low energy) elementary excitations of one state become the (high energy) 
topological excitations of another. The supersymmetric mean-field theory seems to support 
this picture. 
 
Notice that because free energy is an increasing function of $\delta$ when $\delta$ 
is small in SFMFT\cite{sign}, 
there exists an instability region 
where the system is unstable towards phase separation
(the area under dotted line) in mean-field 
theory. 
A possible consequence of this instability is that stripe phase might 
develop in this part of the phase diagram.  
Notice however, that our mean-field free energy is probably not accurate enough to describe the 
instability region correctly.

The first-order phase transition between slave-boson-like and 
slave-fermion-like phases implies that our phase diagram is extremely sensitive to the free energy 
of the two phases. In particular, it is well known that the mean-field free energies are 
unreliable because of the relaxation of constraint 
of no double occupancy. A better way to compute free energies is to use Gutzwiller's 
projected wavefunction \cite{gutz1} $|\phi_G\rangle=P|\phi_o\rangle$, 
where $|\phi_o\rangle$ is a mean-field state in the supersymmetric theory, and 
$P=\Pi_{i}\delta(h_i^+h_i+\sum_{\sigma}\psi^+_{i\sigma}\psi_{i\sigma}-1)$ 
is the projection operator. We shall estimate the energies of the wavefunctions $|\phi_G\rangle$ 
using an approximate method originated by Gutzwiller \cite{gutz1} in the following.
  
Let $A$ be any operator where we want to find its ground state expectation value 
$\langle\phi_G|A|\phi_G\rangle$. Gutzwiller introduces the factor $g_A$ such that\cite{gutz1}
$\langle\phi_G|A|\phi_G\rangle=g_A \langle\phi_o|A|\phi_o\rangle$
where the coefficient $g_A$ is determined approximately by considering only the ratio of 
{\em classical weights} of the Gutzwiller projected state and that of mean-field 
state\cite{gutz1,gutz2}.
Applying the above procedure to our supersymmetric mean-field theory, we find that
there are four kinds of $g$ factors, where two come from Heisenberg interaction term 
in slave-boson and slave-fermion mean-field states, respectively and the other two come from the kinetic 
energy terms. We found that $g_J^{sb}=4/(2-\delta_f)^2$
for the Heisenberg interaction term in slave-boson mean-field state, 
where $\delta_f=\sum_{\sigma}\langle f_{i\sigma}^{\dag}f_{i\sigma}\rangle$ and
$g_J^{sf}=1$ in the slave-fermion mean-field state. Similarly, 
\begin{eqnarray}
g_t^{sb}=\frac{2(1+\delta_{\cal B})}{(2-\delta_f)(\sum_{n=1}^{\infty}\sqrt{n}(\frac{\delta_{\cal B}}{1+\delta_{\cal B}})^n)^2} , 
g_t^{sf}=\frac{2+\delta_b}{2(1-\delta_{\cal F})(\sum_{n=1}^{\infty}\sqrt{n}(\frac{\delta_b}{2+\delta_b})^n)^2} 
\nonumber 
\end{eqnarray}
for the hopping terms, where 
$\delta_b=\sum_{\sigma}\langle b_{i\sigma}^{\dag}b_{i\sigma}\rangle$, 
$\delta_{\cal F}=\langle {\cal F}_i^{\dag}{\cal F}_i\rangle$ and 
$\delta_{\cal B}=\langle {\cal B}_i^{\dag}{\cal B}_i\rangle$. 
The details of these calculations will be reported in a separate paper.

After applying the Gutzwiller's method, the resulting $\delta$-$T$ phase diagram for the same 
value of $t/J$=3 is shown in Fig.2. Comparing with the bare mean-field phase diagram 
we find that the slave-fermion-like spiral phase has shrinked into lower doping regime. 
The transition doping concentration at zero temperature drops from 0.23 to 0.06, 
which shows that 
the importance of constraint in obtaining the correct 
free energies. The d-wave RVB pairing phase also extends to higher temperature
region and covers a larger area in the phase diagram, in agreement with general 
experimental findings. 

In summary, we present in this paper the mean-field phase diagram of $t$-$J$ model 
in supersymmetric representation.
The supersymmetric theory produces several new features not present in usual
slave-boson or slave-fermion treatment of {\it t-J} model. 
Besides providing a better phase diagram compared with experiment, 
new branches of spin and hole excitations are obtained in supersymmetric theory. 
The transition from slave-fermion-like spiral state to slave-boson-like d-wave
superconducting state is first-order in our mean-field treatment, similar to what
is obtained in SO(5) theory. The stability of our mean-field phase diagram against
quantum fluctuations is unknown at present, and shall be a subject of our future works.
Nevertheless, our success in constructing a theory
which incorporates both the advantages of the slave-fermion and slave-boson 
mean-field theories suggests that supersymmetric formulation can be a fruitful starting point 
to understand the relation between antiferromagnetism and superconductivity 
in {\it t-J} model.

We acknowledge support of HKUGC through RGC Grant No. HKUST6143-97P.

\begin{figure}
\caption{Phase Diagram of {\it t-J} model with $t/J$=3. The phase diagram has four
regions. Region I: slave-fermion-like spiral phase. 
Region II: Slave-boson-like d-wave RVB pairing phase.
Region III: Slave-boson-like normal state.
Region IV: High-temperature phase.
The area under the dotted line is unstable to phase separation in mean-field theory.
The dashed line is the bose-condensation temperature for slave-boson-like
solutions.}
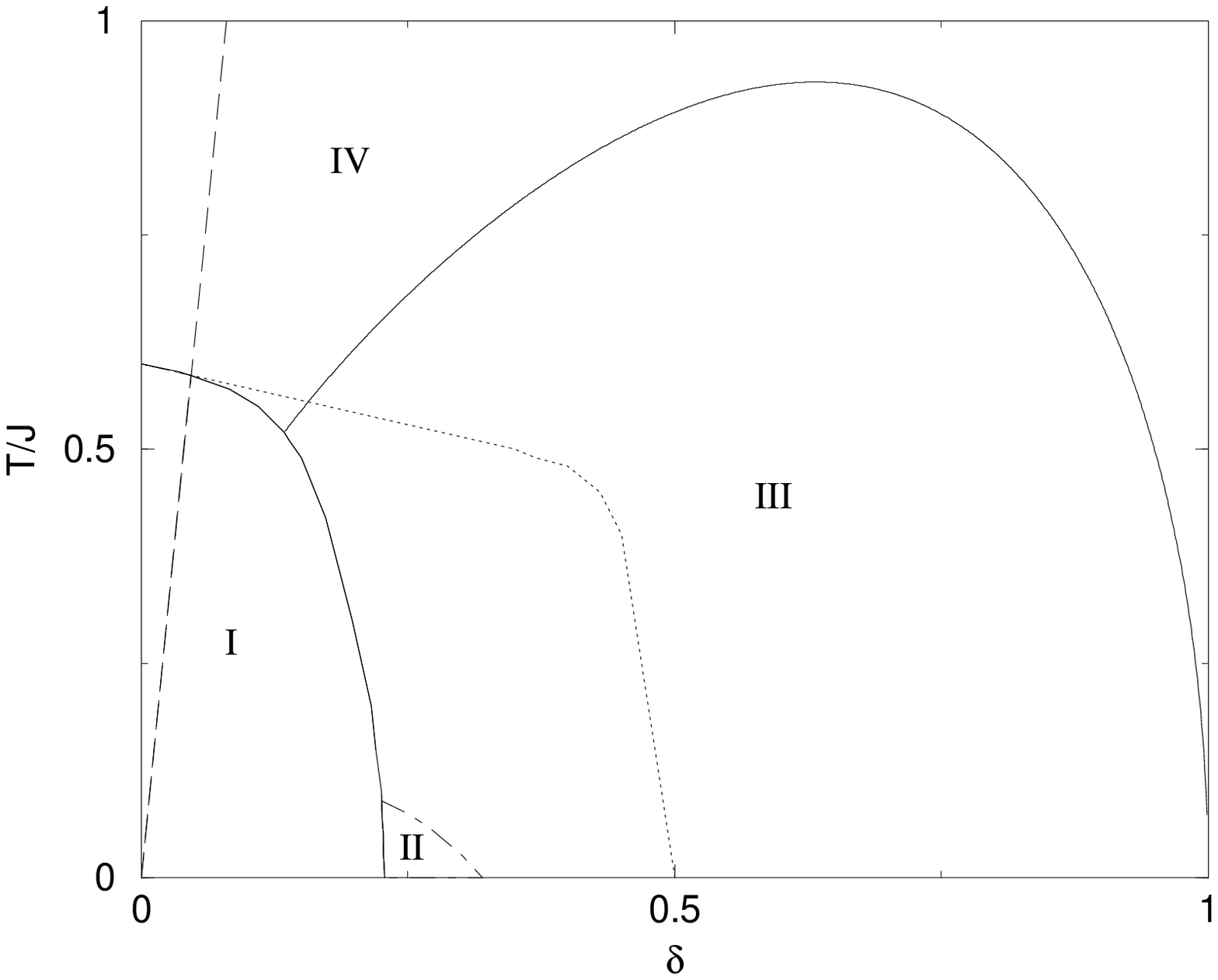
\label{fig1.eps}
\end{figure}

\begin{figure}
\caption{Phase Diagram of $t/J$=3 corrected by Gutzwiller's method.
The labeling is the same as that in Fig.1.} 
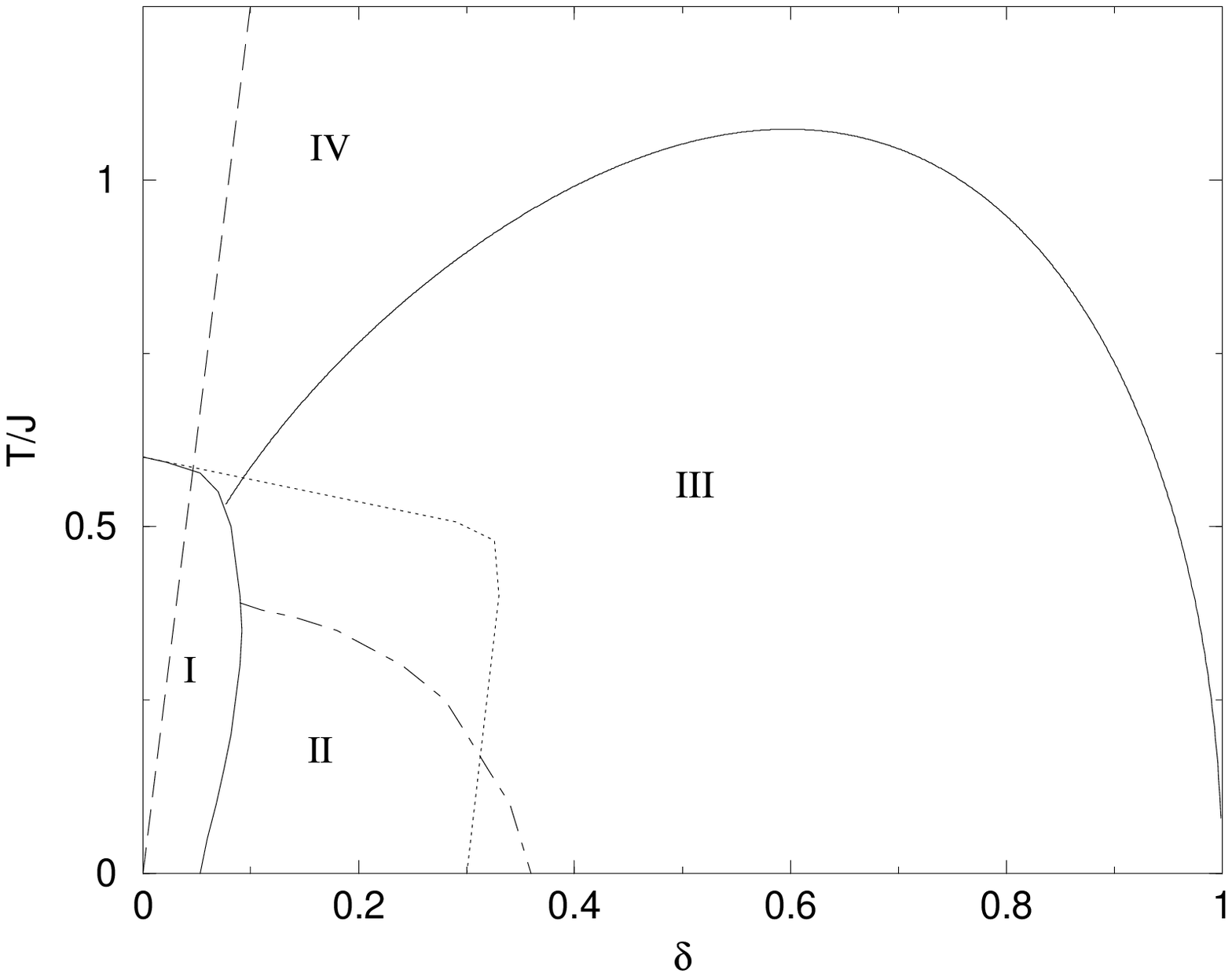
\label{fig2.eps}
\end{figure}


\end{document}